\documentclass [a4paper,twocolumn] {article}

\usepackage[utf8]{inputenc}
\usepackage[english]{babel}

\usepackage{geometry}
\usepackage{graphicx}
\usepackage{physics}
\usepackage{braket}
\usepackage{bm}
\usepackage{float}
\usepackage{siunitx}

\usepackage[affil-it]{authblk}
\usepackage{blindtext}
\usepackage{abstract}
\usepackage{comment}
\usepackage{amsmath}
\usepackage{amssymb}
\usepackage{bbold}

\usepackage{caption}

\usepackage{url}
\DeclareUnicodeCharacter{03B1}{$\alpha$}
\DeclareUnicodeCharacter{03B3}{$\gamma$}

\usepackage{xcolor}
\usepackage{adjustbox}

\usepackage[numbers,sort&compress]{natbib}

\newcommand{\al}{$\alpha$~}
\newcommand{\ga}{$\gamma$~}

\title{A configuration interaction approach to solve the Anderson impurity model; applications to elemental Ce}

\author[1]{B. Herzog}
\author[1]{P. Thunström}
\author[1,2]{O. Eriksson$^{*}$}
\affil[1]{Department of Physics and Astronomy, Uppsala University, Box 516, 751 20 Uppsala, Sweden}
\affil[2]{Wallenberg Initiative Materials Science (WISE), Uppsala University}

\date{}

\begin{document}

\twocolumn[\begin{@twocolumnfalse} 
    
\maketitle 

* Corresponding author, email: \texttt{olle.eriksson@physics.uu.se}

\begin{abstract}
 Accurate calculations of strongly correlated materials remain a formidable challenge in condensed matter physics, particularly due to the computational demand of conventional methods. This paper presents an efficient solver for dynamical mean field theory using configuration interaction (CI). The method is shown to have improved efficiency compared to traditional, exact diagonalization approaches. Hence, it provides an accessible, open-source alternative that can be executed on standard laptop computers or on supercomputers. The solver is demonstrated on cerium in the $\gamma$-, $\alpha$- and $\epsilon$-phases. An analysis of how the electronic structure of Ce evolves as function of lattice compression is made. It is argued that the electronic structure evolves from a localized nature of the 4f shell in $\gamma$-Ce to an essentially itinerant nature of the 4f shell of $\epsilon$-Ce. The transition between these two phases, as function of compression, can hence be seen as a Mott transition. However, this transition is intercepted by the strongly correlated $\alpha$-phase of elemental Ce, for which the 4f shell forms a Kondo singlet.
\vspace{1cm}
\end{abstract}
\end{@twocolumnfalse}]

\section{Introduction}

In the field of condensed matter physics, accurate calculations of electronic interactions in strongly correlated materials remains a significant challenge that is pivotal for unraveling physical and chemical properties of complex systems. The Anderson impurity model, instrumental in studying magnetic impurities, the Kondo effect, and its central role in the dynamical mean field theory (DMFT), has proven indispensable for understanding electron correlation effects in a variety of contexts. This paper introduces a novel impurity solver developed using a configuration interaction (CI) framework, tailored to enhance the computational efficiency and accuracy of such models. It is exemplified by a DMFT calculation of the electronic structure of three allotropes of elemental Ce.

Traditionally, solutions to the Anderson impurity model have employed the numerical renormalization group (NRG), quantum Monte Carlo (QMC) simulations, and exact diagonalization (ED) methods\cite{Paul2019}, each with inherent limitations in terms of scalability, temperature range for applicability, and computational demands. The configuration interaction method offers a controlled approach to solving the Anderson impurity model, by systematically including electron-electron interactions within a truncated Hilbert space. This method's adaptability to various basis set sizes and its ability to provide precise solutions at zero temperature make it a powerful tool for probing ground state and excited state properties. CI belonging to the class of ED methods, it thus inherits its advantages compared to other impurity solvers: real frequency computations of green functions are accessible, without the need for analytical continuation, as in Matsubara frequency-based calculations. In addition, general forms and strengths of the Hamiltonian are unproblematic, e.g. Coulomb interaction and crystal field effects in parallel to a sizable spin-orbit coupling is unproblematic, in contrast to solutions based on continuous-time quantum Monte Carlo (CT-QMC) solvers where the well known sign problem becomes apparent\cite{Kim2020}.\\

Although traditional ED methods are routinely used to study strongly correlated materials, the low number of computationally accessible fermionic degrees of freedom hinders a faithful representation of the hybridization function in the DMFT method.
In recent years, different impurity solvers have emerged that fall within the broad category of "truncated Hilbert space" solvers. The general idea, borrowing from quantum chemical methods dating back to the late 1960s \cite{Whitten1969}, is that not all Slater determinants are equally important, and most of them have negligible weight in a given region of the spectrum. Thus, instead of wasting numerical resources in diagonalizing the problem in the full basis, one should look for optimized truncations of the latter. The potential of Configuration Interaction methods for quantum impurities, DMFT and Hubbard models was already demonstrated by Zgid \emph{\& al.} \cite{Zgid2012} in 2012. A following approach, using the Lanczos algorithm to search for low energies states by explicit application of the Hamiltonian operator, was used by Lu \emph{\& al.} who demonstrated with single site DMFT on the Bethe lattice, that up to 300 bath states were possible to use to represent the hybridization function \cite{lu_efficient_2014,lu_exact_2017,lu_natural-orbital_2019}. 

More recently, machine learning was used to select the most important Slater determinants for a single Anderson Impurity Model by Bilous \emph{\& al} \cite{Bilous2025}. While it is now evident that such approaches are efficient at solving strongly correlated problems, there is still a lack of an open-source implementations dedicated to treat complex multi-orbital impurity problems, integrated within a DFT+DMFT framework, in order to describe realistic materials. In this work, we present such an implementation, and first demonstrate its applicability for the famous $\alpha$ to $\gamma$ transition of elemental Cerium. The $\alpha$ and $\gamma$ phases of Cerium, discussed in Refs. \cite{OE1,OE2,OE3,OE4,OE6,OE7,OE8,OE9,OE10,OE11,OE12,OE13,OE14,OE15,OE16,OE17}, represent a striking example of how strong electronic correlations can modify the properties of a material. At room temperature an isostructural first-order phase transition occurs between a low-density $\gamma$-phase and a high-density $\alpha$-phase, as function of hydrostatic pressure. Both phases share the same face-centered cubic (fcc) structure, but the $\alpha$-phase has a ~15\% smaller volume compared to the $\gamma$-phase. Furthermore, $\alpha$-Ce has a non-magnetic 4f shell, in contrast to the free ionic value of the magnetic moment of $\gamma$-Ce \cite{OE3}. In the pressure-temperature phase diagram of Ce, the $\alpha$-phase is found to be stable at ambient pressure and temperatures $\lesssim 300K$. At elevated temperatures $\gtrsim 300K$ and finite pressure the $\alpha$-phase can also be stabilized, e.g. at room temperature a pressure of $ \sim 1 GPa$ stabilizes this phase  \cite{OE3}. The transition between $\gamma$- and $\alpha$-Ce is generally believed to be driven by the interplay of electron correlation effects, often described in terms of a Mott transition  \cite{OE1} or a Kondo volume collapse (KVC)  \cite{OE2}, where the competition between 4f electron localization and hybridization with conduction electrons governs the phase behavior. This phase transition, accompanied by significant volume collapse and ending in a critical point at high temperatures, serves as a classic example of correlated electron physics. In fact, several DFT+DMFT calcuations have been published for these two phases of Ce  \cite{OE8,OE9,OE10,OE11,OE12,OE13}.

Several high-pressure phases have been reported for Ce, e.g. the $\alpha^{\prime}$- (orthorhombic), $\alpha^{\prime \prime}$- (monoclinic) and $\epsilon$-phases (body centered tetragonal-bct), as discussed in e.g. Ref.\cite{OE16}. At room temperature the $\alpha^{\prime}$-phase, or a two phase region with $\alpha^{\prime}$ and $\alpha^{\prime \prime}$ is stable in the pressure range 4-12 GPa (see Fig.1 of Ref.\cite{OE16}). At higher pressures the $\epsilon$-phase is reported to be stable. Interestingly at even higher pressures, Ref.\cite{OE16} predicts a re-entrance of the fcc structure, in what was named the $\omega$-phase. 

The low symmetry phases of Ce have been discussed to reflect a rather large degree of itinerancy of the 4f shell \cite{OE14,OE16}, something which the high pressure phases of Ce share with the actinide elements; Pa, U, Np and Pu all exhibit low symmetry crystal structures \cite{OE18}. The argument put forth in Refs.\cite{OE14,OE18,OE16} is that narrow energy bands, provided by the electronic f-shell, can undergo a Peierls distortion that stabilize the observed low symmetry structures discussed above. However, narrow energy bands are also archetypical systems where electronic multi-configurations (correlations) become important, and for this reason we have included in this study also the $\epsilon$-phase. It should be noted that the electronic structure of the ambient- and high-pressure phases of Ce have been investigated by DMFT methods, particularly with a continuous-time quantum Monte Carlo
impurity solver. These results are reported in Refs.\cite{OE6,OE13,OE17} and we will compare our results to previous data.

\section{Results}
\subsection{$\alpha$- and $\gamma$-cerium }

\begin{figure}[H]
    \centering
    \hspace*{-0.5cm}
    \includegraphics[width=0.5\textwidth]{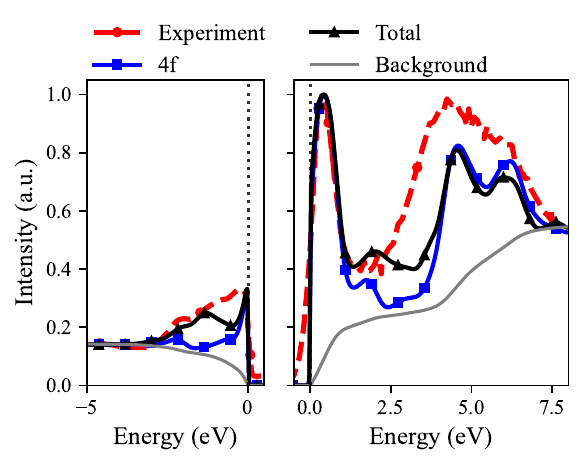}
    \caption{Experimental and theoretical spectral function of $\alpha-$Ce for occupied states (left figure) and unoccupied states (right figure). Experimental data are from Ref.\cite{wieliczka_high-resolution_1984}. The intensity difference between unoccupied and occupied states was chosen to represent the occupation of available states.}
    \label{fig:alpha_dos}
\end{figure}

\begin{figure}[H]
    \centering
    \hspace*{-0.5cm}
    \includegraphics[width=0.5\textwidth]{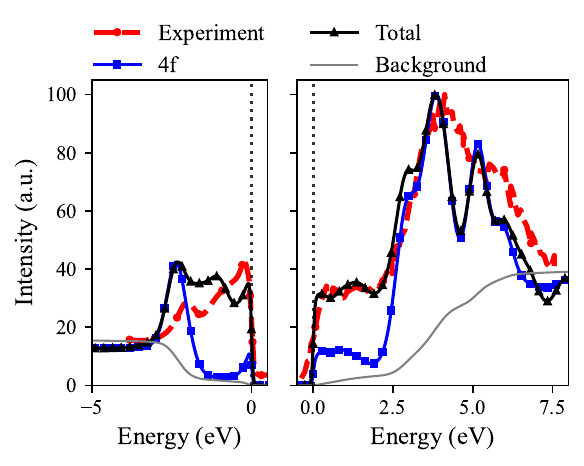}
    \caption{Experimental and theoretical spectral function of $\gamma-$Ce for occupied states (left figure) and unoccupied states (right figure). Experimental data are from Ref.\cite{wuilloud_electronic_1983}. The intensity difference between unoccupied and occupied states was chosen to represent the occupation of available states.}
    \label{fig:gamma_dos}
\end{figure}

In Figures \ref{fig:alpha_dos} and \ref{fig:gamma_dos} we compare the computed and experimental spectral functions for the \al and \ga phases of cerium, respectively. Experimental occupied and unoccupied spectra are from x-ray photoelectron spectroscopy\cite{wieliczka_high-resolution_1984} (the \SI{40}{eV} incident photon energy curves) and bremsstrahlung isochromat spectroscopy (BIS)\cite{wuilloud_electronic_1983}, respectively.
We show both the $4f$ projected density of states, as well as the total density of states: at \SI{40}{eV}, the $5d$ and $4f$ extrapolated photoemission cross sections match \cite{vandereb-2000}, and for the BIS spectra, there is no available cross section data to use.
A background was added in the form of a cumulative function to match the intensities of the experiments at a binding energy below the bottom of the valence band. The height of the most prominent peak was used to connect the theoretical and experimental spectra. A Fermi-Dirac convolution is applied with a temperature used for each phase of Ce as specified in the section on computational details. An additional gaussian broadening of 0.1 eV is added.

The most natural way to assess how a DMFT solver performs, is to compare how it captures key features of a measured spectral function, including quasiparticle peaks, satellite features, and relative intensity variations between the several phases, if possible. For Ce such data are fortunately available, and in Figure \ref{fig:alpha_dos} one may observe that the calculated spectral function of $\alpha$-Ce exhibits a sharp Ce $4f$ quasiparticle peak just above the Fermi level, in agreement with BIS experimental data \cite{wuilloud_electronic_1983}. This peak is associated with the Kondo resonance, which characterizes the many-body nature of the $4f$ electrons in $\alpha$-Ce. In addition, two features at \SI{-2.2}{\eV} and \SI{4.5}{\eV} are present, corresponding to the lower and upper Hubbard bands. The corresponding gap of \SI{6.7}{\eV} slightly overestimates the experimental gap of \SI{6.4}{\eV}. We also note that from theory, the impurity occupation is $n^{\alpha}_f = 0.88$, 7\% smaller than the experimental estimates ($n^{\alpha}_f(\text{Exp.)} = 0.95$ \cite{RUeff2006}). 


The results of $\gamma$-Ce, shown in Figure \ref{fig:gamma_dos}, exhibits a significantly reduced Ce $4f$ quasiparticle peak, indicating stronger localization and less hybridization of 4f electrons. The calculated gap between the Hubbard bands, \SI{6.2}{\eV}, matches that of the experimental spectra, and the intensity of spectral features close to the Fermi level are in good agreement with experiment as well. The low hybridization of $\gamma$-Ce is in agreement with general discussions on the properties of $\gamma$-Ce \cite{OE3} and the fact that the spectrum shown in Figure \ref{fig:gamma_dos} agrees quite well with results of DMFT calculations that employ the non-hybridizing Hubbard-I approximation for the impurity solver \cite{OE4}.
From the theoretical calculations of $\gamma$-Ce we find a 4$f$ occupancy of $n^{\gamma}_f = 0.99$, which agrees well with the experimental value of $n^{\gamma}_f(\text{Exp.)} = 0.97$ \cite{RUeff2006}.

\begin{figure}[H]
    \centering
    \includegraphics[width=0.5  \textwidth]{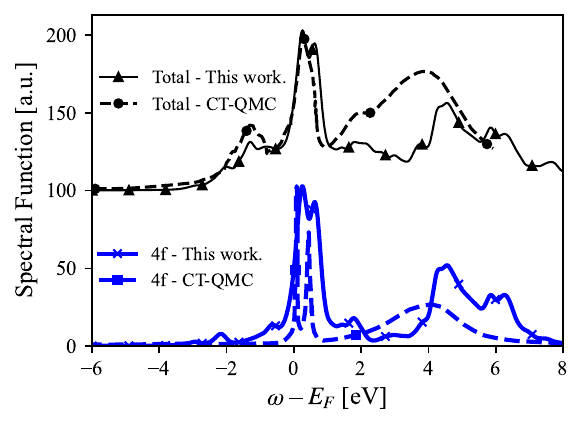}
    \caption{$\alpha-$Ce: Comparison of CT-QMC total spectral function \cite{OE10} with this work (upper panel) and the $4f$ projected results (lower panel). For details of the calculations see main text.}
    \label{fig:ctqmc_alpha}
\end{figure}

\begin{figure}[H]
    \centering
    \includegraphics[width=0.5  \textwidth]{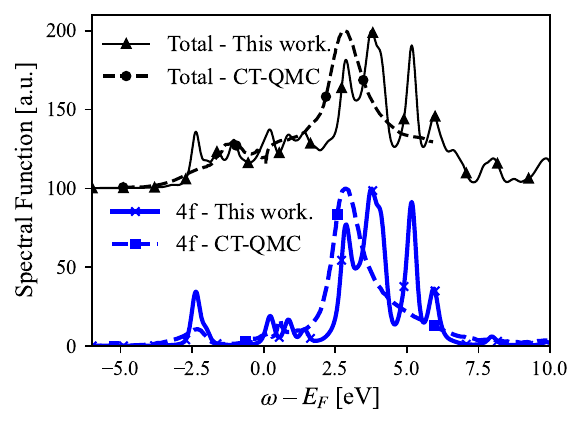}
    \caption{$\gamma-$Ce: Comparison of CT-QMC total spectral function \cite{OE10} with this work (upper panel) and the $4f$ projected results (lower panel). For details of the calculations see main text.}
    \label{fig:ctqmc_gamma}
\end{figure}

In Figs. \ref{fig:ctqmc_alpha} and \ref{fig:ctqmc_gamma} we compare for $\alpha$- and $\gamma$-Ce our bare $4f$ and total (including $spd$ contribution) spectral functions with spectral functions computed using charge self-consistent CT-QMC calculations\cite{OE10}: the later were obtained using the CT-QMC within the hybridization expansion method. Values for $U$ and $J$ were 6 eV and 0.6 eV respectively, and a spin-orbit coupling Hamiltonian with parameter $\lambda_{so}=0.0953$ eV was used. The spectra were normalized to the same maximum peak height for easier comparison. There is an overall good qualitative agreement between both approaches. For the $\alpha$ phase, CT-QMC shows a much stronger Kondo peak at the Fermi level, while the lower Hubbard band is barely visible in the $4f$ spectral function. In the $\gamma$ phase, CT-QMC underestimates the position and width of the higher Hubbard band. 

\begin{figure}[H]
    \centering
    \includegraphics[width=0.5  \textwidth]{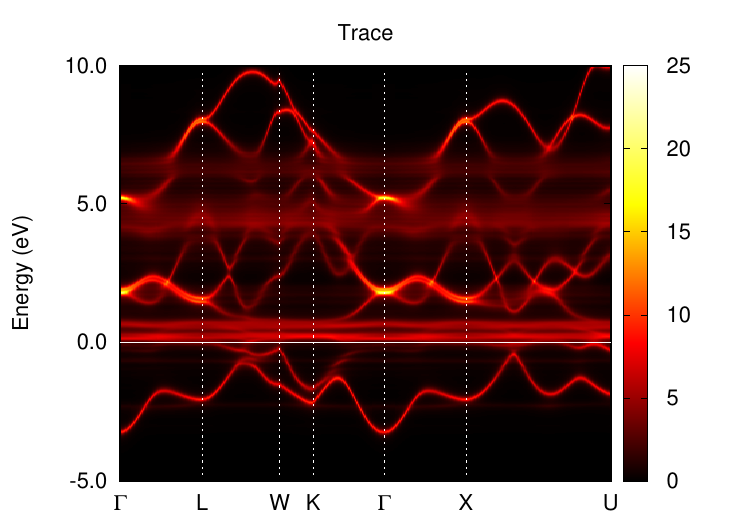}
    \caption{Energy dispersion of $\alpha-$Ce obtained from the same calculation as shown in Fig.\ref{fig:alpha_dos} for k-integrated data.}
    \label{fig:alpha_bands}
\end{figure}

\begin{figure}[H]
    \centering
    \includegraphics[width=0.5  \textwidth]{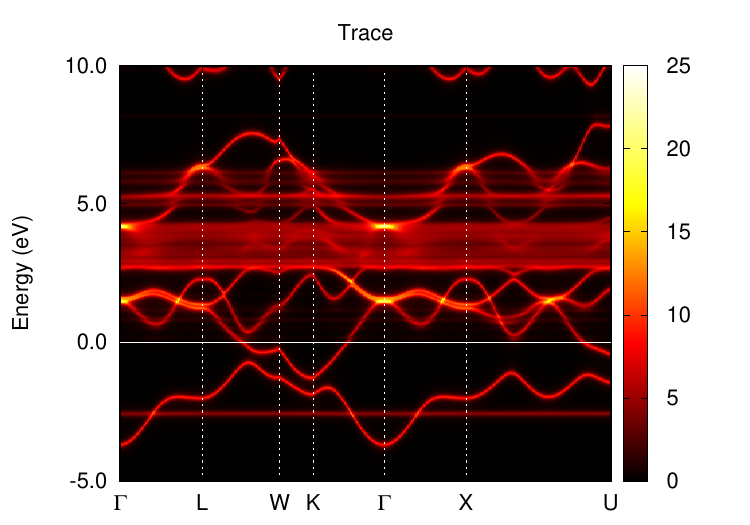}
    \caption{Energy dispersion of $\gamma-$Ce obtained from the same calculation as shown in Fig.\ref{fig:gamma_dos} for k-integrated data.}
    \label{fig:gamma_bands}
\end{figure}

In order to facilitate a comparison between theory and experiment for the dispersion of the electronic structure, we show the k-resolved spectral function of $\alpha$- and $\gamma$-Ce in Figs. \ref{fig:alpha_bands} and \ref{fig:gamma_bands}, respectively. For both phases coherent quasiparticle states are apparent, where a well-defined connection between crystal momentum and energy are observed essentially for all occupied states. For states at, and just above, the Fermi level, $\alpha$-Ce displays a more complicated relationship between crystal momentum and energy. Instead of sharp features in the energy dispersion, correlation-driven life-time effects broaden the energy states so that a pure energy band structure is not discernible. For $\gamma$-Ce this is not the case, since the reduced hybridization leads to an absence of $4f$ spectral weight at, and just above, the Fermi energy. As a result the electronic structure of $\gamma$-Ce shows clear Ce $5d$ quasiparticle behavior also at, and just above, the Fermi energy. For energies corresponding to the upper Hubbard band, both phases show clear deviation from a pure quasiparticle picture, with significant life-time broadening of non-dispersive $4f$ states.

\begin{figure}[H]
    \centering
    \hspace*{-0.5cm}
    \includegraphics[width=0.5\textwidth]{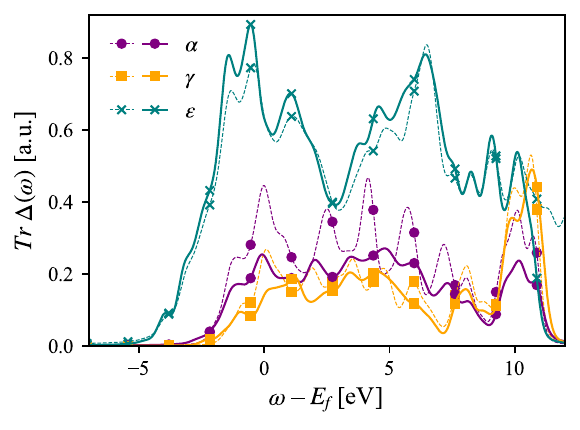}
    \caption{Trace of the hybridization functions with respect to the $4f$ shell for $\alpha$, $\gamma$ and $\epsilon$ cerium, from converged DFT (dashed lines) and DMFT (solid lines) calculations.}
    \label{fig:hyb}
\end{figure}

The crystal structure of $\alpha$- and $\gamma-$Ce are the same; fcc, with the only difference being their volumes. One may ask then what the main difference is between these two phases, that causes the very large difference in electronic structure discussed above. According to Eqn. \eqref{SIAM}, there are in fact not that many terms to track down to explain this marked difference. As outlined in the section on computational details, we have used rather similar values of Hubbard U for these two phases, which leaves the hybridization between $4f$-orbitals and the DMFT bath states as the most likely cause of difference (also discussed in Ref.\cite{herper_combining_2017}). To illustrate this point, we show 
in Fig. \ref{fig:hyb} the hybridization function, computed at DFT level. It can be noted that the general features of $\alpha$- and $\gamma-$Ce are similar, but importantly that $\alpha-$Ce has about twice as high values of the hybridization function, compared to that of $\gamma-$Ce. This subtle scaling difference, particularly at the Fermi level, is responsible for the dramatic difference in the $4f$ spectral function, as the more hybridized state has significantly larger coupling between the $4f$-levels and the purely itinerant electron states.

\subsection{$\epsilon$-cerium }

To further showcase the versatility of our CI solver and to explore the electronic structure of cerium under more extreme conditions, we turn our attention to the $\epsilon$ phase. Stable at high pressures (typically for $P \gtrsim 12$ GPa and at a volume $V \approx$ \SI{20}{\angstrom} at room temperature), $\epsilon-$Ce crystallizes in a body-centered tetragonal (bct) structure. The reduced volume causes an increase in the wavefunction overlap and hybridization of the $4f$ states, with an increased bandwidth as a result. This modifies the balance between kinematic effects when electrons travel around in the lattice and Coulomb repulsion when two $4f$ electrons are found on the same atomic site, as specified by Eqn. \eqref{SIAM}. Increased wavefunction overlap tends to favor kinematic effects over Coulomb repulsion, causing band like electron states. We note that the bct phase has in fact been suggested to be driven by essentially itinerant, but narrow, electron states, that tends to form low symmetry structures of Ce and for some of the actinides \cite{OE14,OE18,OE16}. Our calculations for $\epsilon-$Ce aim to characterize the balance between band formation and Coulomb repulsion, as signaled by spectral features, and to assess the degree of $4f$ electron correlation in this high-density environment.

\begin{figure}[H]
    \centering
    \includegraphics[width=0.5  \textwidth]{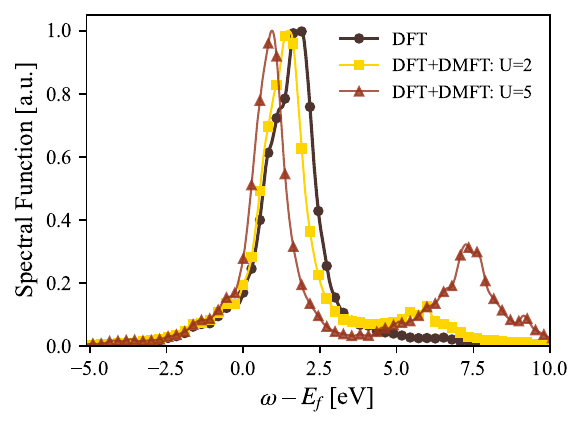}
    \caption{Projected $4f$ spectral function for $\epsilon-$Ce, using DFT and DFT+DMFT with $U=\SI{2}{\eV}$ and $U=\SI{5}{\eV}$. For details see main text.}
    \label{fig:epsilon_dos}
\end{figure}

\begin{figure}[H]
    \centering
    \includegraphics[width=0.5  \textwidth]{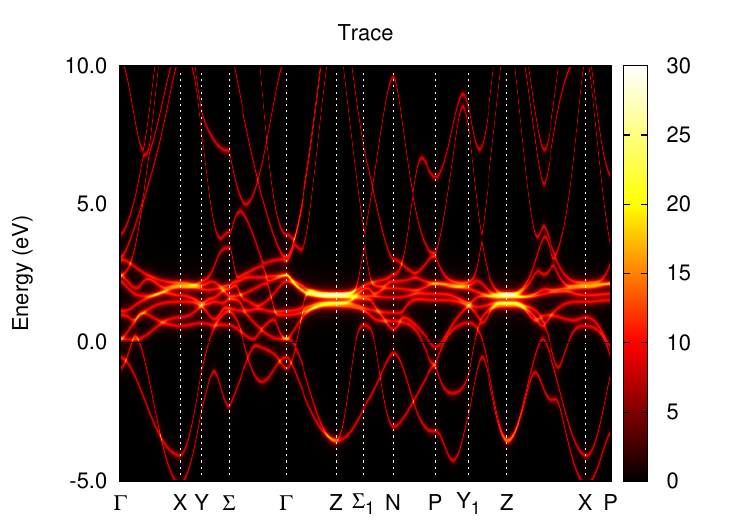}
    \caption{Energy dispersion of $\epsilon-$Ce obtained from the same DFT calculation as shown in Fig.\ref{fig:epsilon_dos} for k-integrated data.}
    \label{fig:eps_band_dft}
\end{figure}

\begin{figure}[H]
    \centering
    \includegraphics[width=0.5  \textwidth]{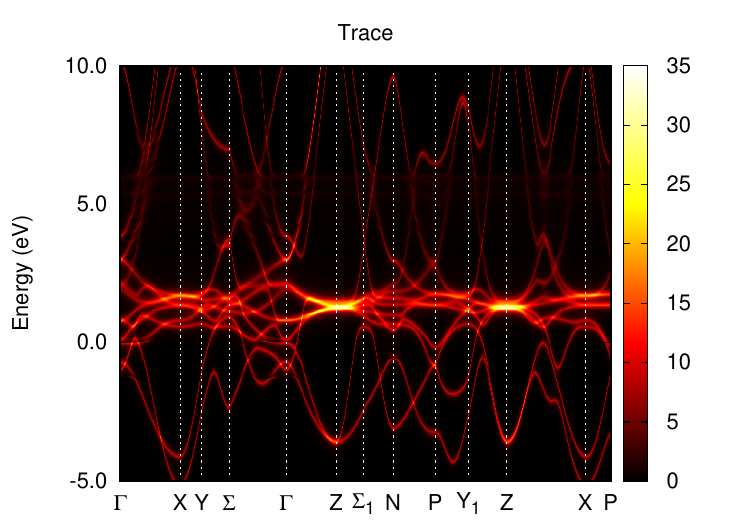}
    \caption{Energy dispersion of $\epsilon-$Ce obtained from the same DFT+DMFT calculation as shown in Fig.\ref{fig:epsilon_dos} for k-integrated data, with U=2 eV.}
    \label{fig:eps_band_u2}
\end{figure}

We used two values of Coulomb repulsion, $U=$ \SI{2}{\eV} and $U=$ \SI{5}{\eV}, for the calculation of $\epsilon$-Ce, while keeping the same $J=$ \SI{0.6}{\eV} as in the calculations of $\alpha-$Ce and $\gamma-$Ce. Due to the lower volume of the $\epsilon-$Ce cell, we expect a higher delocalization of the $4f$ shell, with an increasing screening and a lower value of the Coulomb repulsion. Figure \ref{fig:epsilon_dos} shows the calculated $4f$-projected spectral function for $\epsilon$-Ce from DFT and DFT+DMFT level of theory. One may note that inclusion of a finite Hubbard U in the calculations causes a narrowing of the electron states, which is according to expectations. However, the spectral features of $\epsilon$-cerium are very different from those of $\alpha$- and $\gamma$-cerium, shown in Figs.\ref{fig:ctqmc_alpha} and \ref{fig:ctqmc_gamma}, respectively. Instead of showing many-body aspects of the spectral function, the inclusion of a Hubbard U for the electronic structure of $\epsilon$-Ce only causes minor deviations from a band behavior, especially for calculations based on $U=$ \SI{2}{\eV} (that given the low volume of this phase, seem to represent the most realistic calculation of the DFT+DMFT data shown in Fig.\ref{fig:epsilon_dos}). The band-like behavior of the DFT+DMFT calculation are also clear when comparing the k-resolved spectral properties of the DFT and DFT+DMFT results, shown in Fig.\ref{fig:eps_band_dft} and \ref{fig:eps_band_u2}, respectively. Apart from life-time broadening of the DFT+DMFT calculation, the k-resolved spectral properties are quite similar in the two calculations, pointing to an electronic structure that is rather close to that given by a DFT level of theory. 

\section{Discussion}

The electronic structure of three different phases of Ce have been analyzed in this study, that of $\gamma$-, $\alpha$- and $\epsilon$-Ce, using dynamical mean field theory coupled to electronic structure methods. In this effort a new impurity solver has been implemented and used. Our results show that the electronic structure of high-volume $\gamma$-Ce is close to being described as a Mott insulator having a fully localized $4f$ shell (bad metal), while that of low volume $\epsilon$-Ce is close to a band-like picture, as given by electronic structure (DFT type) calculations. The intermediate volume ($\alpha$) phase, has a more complex electronic structure, with signatures of many-body physics, like an upper and lower Hubbard feature and a Kondo like peak at the Fermi level. 

From these results we conclude that the much discussed Mott transition in Ce \cite{OE1} happens between the high volume $\gamma$-phase and the low volume $\epsilon$-phase (or possibly to the $\alpha^{\prime}$- or  $\alpha^{\prime \prime}$-phase), as opposed to the previously discussed transition between $\gamma$- and $\alpha$-Ce. The $\alpha$-phase disrupts a pure Mott transition in Ce, as it interjects the transition between fully localized and fully itinerant states, with an intermediate, but highly interesting, electronic structure that is heavily influenced by a competition between kinematic effects, from the itinerancy of the electrons, and the Coulomb repulsion that appears when two $4f$ electrons are on the same atomic site.

It is instructive to compare the performance of the here discussed CI-based solver, where machine learning algorithms are used to accelerate the identification of relevant configurations, with the widely used continuous-time quantum Monte Carlo (CT-QMC) method, as these approaches are best viewed as complementary tools with distinct domains of applicability.
The primary advantage of the CI-based solver is its ability to compute the Green's function directly on the real-frequency axis, thereby circumventing the numerically ill-posed problem of analytic continuation that is required for CT-QMC. Furthermore, as a deterministic, wavefunction-based method, a CI-based approach is immune to the fermion sign problem. This allows for a robust treatment of complex, multi-orbital Hamiltonians with strong spin-orbit coupling or general crystal field terms, which often render CT-QMC simulations computationally intractable. 

A testament to the remarkable efficiency of the CI-based approach is that all calculations for this study were performed on a standard laptop, with a single DMFT iteration typically completing in 10 minutes. Conversely, the main limitation of any ED-based solver is the discretization of the bath, which restricts its low-energy resolution. CT-QMC is therefore superior for resolving the exponentially narrow Kondo resonances found in true local-moment systems at low temperatures. The success of the CI-based method for $\alpha$-Ce stems precisely from the fact that it is an intermediate valence system with a very high Kondo scale (~1000 K), resulting in a broad spectral feature at the Fermi level that is well-resolved by our discrete bath. Thus, for a broad class of correlated materials where the sign problem is severe or where an unbiased view of the real-frequency spectral function is paramount, the CI approach presented here represents a highly efficient and powerful alternative. 

In terms of actual electronic structures, we note that for the $\alpha$-phase CT-QMC and the CI-approach suggested here, give rather similar results (see Fig.4). The main difference is that the CI-approach results in a more pronounced peak associated to the lower Hubbard band as well as more distinct spectral features at higher energies. For the $\gamma$-phase the lower Hubbard band is also more pronounced with the CI-approach, compared to CT-QMC, and multiplet features appear more naturally in the unoccupied states.  

\section{Methods}

\subsection{Impurity Hamiltonian}

The impurity Hamiltonian, which is a central aspect of dynamical mean field theory coupled to density functional theory \cite{DMFT1}, is obtained with the electronic structure method outlined in Refs.\cite{DMFT2,DMFT3,DMFT4,Sorgenfrei2024}, and briefly summarized here. It reads 
\begin{equation}
    \hat H = \hat h + \hat U_{imp}
    \label{SIAM}
\end{equation}

\noindent
$\hat U_{imp}$ is a two-particle rotationally invariant interaction acting on the impurity degrees of freedom (here the Cerium $4f$ orbital shell), parametrized by two parameters $U$ and $J$.

\begin{equation}
    \hat h = \begin{pmatrix}
    \hat h_{imp} & \hat V^{\dagger} \\ \hat V & \hat h_{bath}
\end{pmatrix}
\label{one-part-ham}
\end{equation}

\noindent
is the one-particle hamiltonian, whose one-particle basis states are separated into impurity and bath subspace. This hamiltonian is obtained from a local DFT hamiltonian $\bm h^{DFT} = \frac{1}{N_k} \sum_k \bm h^{DFT}_k$ with $N_k$ the number of $k-$points sampled in the Brillouin zone and $\bm h_k^{DFT}$ the Kohn-Sham DFT hamiltonian. This Hamiltonian is projected onto a local basis set that separates correlated orbitals (the impurity subspace) from the rest. We denote these blocks as $\bm h^{DFT}_{imp}$, $\bm h^{DFT}_{r}$, and their couplings $\bm h^{DFT}_{imp-r}, \bm h^{DFT}_{r-imp}$.

The local non-interacting green function $\bm G^0(z)$ is then given by 
\begin{align}
    \bm G^0(z) &= (z\mathbb 1 - \bm h^{DFT})^{-1} \\ \nonumber
    &= \begin{pmatrix} z\mathbb 1 - \bm h_{imp}^{DFT} & - \bm h_{imp-r}^{DFT} \\ - \bm h_{r-imp}^{DFT} & z\mathbb 1 - \bm h_{r}^{DFT}\end{pmatrix}^{-1}
\end{align} so that, using the block inversion formula, the upper left element is

\begin{equation}
\begin{split} 
\bm G_{imp}^0(z) = \\ ((z\mathbb 1 - \bm h_{imp}^{DFT}) -
\bm h_{imp-r}^{DFT}(z\mathbb 1 - \bm h_r^{DFT})^{-1}\bm h_{r-imp}^{DFT})^{-1} \\
= (z\mathbb 1 - \bm h_{imp}^{DFT} - \bm \Delta(z))^{-1}
\end{split}
\end{equation}

The last equality defines the hybridization function $\bm \Delta(z)$. The latter is further discretized into a finite set of $n_b$ poles per element $\Delta_{ij}(z) \approx \sum_{\alpha=1}^{n_b}V^{\dagger}_{i\alpha}(z\ - h^{bath}_{\alpha \alpha})^{-1} V_{\alpha j}= {\tilde \Delta_{ij}}(z)$, or
\begin{equation}
    \tilde{\bm \Delta}(z) = \bm V^{\dagger} (z\mathbb 1 - \bm h_{bath})^{-1} \bm V
    \label{hyb_approx}
\end{equation}

To obtain the approximate hybridization $\bm{\tilde\Delta}$, we minimize the cost function 
\begin{equation}
    \chi = \sum_{\omega}\frac{1}{|\omega + i\eta|}||\bm \Delta(\omega+i\eta) - {\bm{\tilde \Delta}} (\omega+i\eta)||_2
    \label{fit}
\end{equation}

We use a different distance to the real axis in the fitting and in the Green functions calculations, with a bigger value for $\eta$ facilitating the self-consistent convergence of the calculations. The Coulomb interaction already accounted for at DFT level is subtracted from the impurity one-particle hamiltonian $\bm h_{imp}$ in the form of a double counting (DC) hamiltonian $\bm h^{DC}$. For a single angular momentum shell impurity, e.g. the $4f$ orbital of cerium, within an unpolarized DFT calculation, and within the Fully Localized Limit (FLL) scheme, the double counting potential is approximated as $\bm h^{DC} = V^{DC} \mathbb{1}$, $V^{DC} = U (n-\frac{1}{2}) - \frac{J}{2}(n-1)$ with $n$ the impurity occupation. Following Ref. \cite{Haule2015}, we use here the nominal double counting scheme where $n$ is the nominal valence occupancy ($n = 1$ for cerium).
Thus $\hat h_{imp} = \hat{h}_{imp}^{DFT} - \hat h^{DC}$ and $\hat V, \hat h_{bath}$ from \eqref{hyb_approx} define the one-particle hamiltonian in \eqref{one-part-ham}.

\subsection{Configuration Interaction}

Configuration Interaction (CI), just as Exact Diagonalization (ED), is a wavefunction method, in which a many-body Hamiltonian is explicitly constructed and diagonalized. The basis of many-body states (configurations) consists of anti-symmetrized products of one-particle functions (Slater determinants). The original idea of CI was to define a suitable one-particle basis as a starting point, such that truncated subspaces of the full CI space remain both computationally tractable and precise. Traditionally, the molecular orbitals obtained through a self-consistent mean-field procedure (Hartree-Fock (HF)) are used as one-particle orbitals. Indeed, the later yield the best single Slater determinant approximation to the true ground-state wavefunction. Subsequent truncations are obtained by substituting occupied orbitals by unoccupied ones in the HF configuration, and a given state $\ket{\Psi}$ is expressed as 
\begin{equation}
\begin{split}
    \ket{\Psi} = c_0 \ket{\Phi} + \sum_{a,r} c_a^r \hat S_a^r \ket{\Phi} + \sum_{\substack{a<b \\r<s}} c_{ab}^{rs} \hat D_{ab}^{rs} \ket{\Phi} \\+ \sum_{\substack{a<b<c \\ r < s < t}} c_{abc}^{rst} \hat T_{abc}^{rst} \ket{\Phi} + \cdots
\end{split}
\end{equation}
where $\ket{\Phi}$ is the Hartree-Fock Slater determinant, and $\hat S_a^r$,$\hat D_{ab}^{rs}$,$\hat T_{abc}^{rst}$, are respectively single, double and triple substitutions operators, exchanging occupied orbital(s) $a,(b,c)$ with unoccupied orbitals $r,(s,t)$, and $c_{a(bc)}^{r(st)}$ represent the expansion coefficient of each configuration.

For example, the set of all single and double substitutions produces the CI Single-Double (CISD) space. As the Hamiltonian consists of one-particle and two-particle terms, two basis elements $\ket{b_1}$ and $\ket{b_2}$ can have a corresponding non-zero Hamiltonian matrix element only if they differ by at most two occupied one-particle orbitals. The Slater-Condon rules are the corresponding formulas to evaluate the matrix elements of the Hamiltonian in the three following cases: $\ket{b_2} = \ket{b1}$, $\ket{b_2} = \hat S_a^r \ket{b_1}$ and $\ket{b_2} = \hat D_{ab}^{rs} \ket{b_1}$. In order to solve the problem in a given truncated basis of slater determinants $\{\ket{b_i}\}_{i=1..N_{det}}$, one must therefore, for each element $\ket{b_i}$, find the corresponding connected basis elements, before one evaluates the matrix elements. The number of such connected elements correspond to the number of monomials in the second quantized Hamiltonian, which typically scales as $\mathcal{O}(K^4)$, where $K$ is the number of spin-orbitals, so that the construction of the Hamiltonian matrix scales as $\mathcal{O}(N_{det} K^4)$ where $N_{det}$ is the size of the basis.

Once the Hamiltonian matrix is constructed, one can diagonalize for the lowest eigenstates using specialized methods such as the Davidson or Lanczos algorithms. 
The process of finding eigenvalues and eigenstates can be done iteratively in the so-called selective CI (SCI) algorithms, that starts from some truncated basis that is used to find the lowest set of eigenstates. The basis elements which contribute little, i.e. below a cut-off value, to this subspace of eigenstates are removed from the basis. A subspace of the connected basis is then generated and added to the main basis. The process is iterated until convergence is obtained for the eigenenergies. The different SCI methodologies differ mainly in the way they select the new basis elements. This can be done by selecting the full connected subspace (all new basis elements obtained by applying the Hamiltonien on the pruned wavefunction), or only a portion of it, either perturbatively or stochastically. In the present work, a combination of perturbation theory and stochastic generation is used. In the latter, instead of selecting new connected determinants with a uniform probability distribution, we employ an importance sampling methodology, in the form of a generative machine learning model \cite{Herzog2023}. Compared to FCI/ED, for the low-energy subspace, this means that a much smaller many-body spaces are needed to achieve the same level of accuracy.

\subsection{Sampling basis states with CIPSI}

The Conﬁguration Interaction using a Perturbative Selection made Iteratively (CIPSI) algorithm \cite{Huron1973,Garniron2019} uses perturbation theory to evaluate important connected configurations: given some wave function $\ket{\psi^0}$ with energy $E^0$, obtained from a calculation with a specific, original  basis, the basis states $\ket{\phi_I}$ not included in $\ket{\psi^0}$ and connected to it are obtained by application of the Hamiltonian on $\ket{\psi^0}$. A perturbation estimate of their coefficient can be obtained from the expression 
\begin{equation}
    c_I = \frac{|\langle \phi_I | H | \psi^0 \rangle|^2}{E^0 - \langle \phi_I |H|\phi_I \rangle}.
    \label{PT2}
\end{equation}
One can therefore augment the original basis by selecting the most important states according to this perturbation scheme. Once a new basis is obtained, the latter is pruned by removing the states whose coefficients in the states of interest are below some threshold, and the process is repeated.
In practice, we choose a threshold of $10^{-12}$ for the squared coefficients, and we double the size of the current basis at each iteration.
As the number of connected states accessible through the Hamiltonian will grow exponentially, at some point computing those connected states will become the bottleneck of the calculation, so that we switch to a stochastic generation for the basis. The details of this scheme are presented in Ref. \cite{Herzog2023} and the most salient features of it are described in the section below. 
\subsection{Sampling basis states with a generative machine learning model}

A restricted Boltzmann machine is defined as a two-layer neural network consisting of one visible layer of $D$ binary units $\bm v$, one hidden layer of $P$ binary units $\bm h$, a weight matrix $\bm W$ of $D \times P$ values connecting those, plus two bias vectors $\bm{a}$ and $\bm b$. A so-called energy for this network is defined as: $E(\bm v, \bm h, \Lambda) = -(\bm a^T \bm v + \bm b^T \bm h + \bm v^T \bm W \bm h)$, where $\Lambda = \{ \bm a, \bm b, \bm W \}$. One also needs a Boltzmann-like probability distribution obtained from: $P(\bm v) = \sum_{\{\bm h\}} e^{-\xi E} / \sum_{\{\bm h, \bm v\}} e^{-\xi E}$ for a given model $\Lambda$. 
The idea is to adjust $\Lambda$ such that $P(\bm v)$ reproduces a probability distribution of some training data, given by a set of vectors $\{ \bm v_i \}$. Once this is done, one can sample a vector $\bm v$ by using the model probability distribution. For the present case, $\bm v$ represents a Slater determinant, and the training data is constructed by sampling determinants using their squared coefficients in the low energy subspace. It should be noted here that at each iteration, while the lowest $n_{low}$ eigenstates are computed from the current subspace, in both the CIPSI or the generative scheme, the basis istelf is expanded by only considering the (possibly degenerate) ground state(s) (using the latter as reference for the perturbation in Eqn. \eqref{PT2}, or to construct the training dataset in the generative machine learning setup.

\subsection{Green function from Lanczos algorithm}

Once the $n_{low}$ lowest eigenstates of the impurity Hamiltonian in Eqn.~\eqref{SIAM} have been obtained, the green function needs to be evaluated in order to obtain the DMFT self-energy. We start by considering the 
retarded green function, with elements $G^{(n)}_{ab}(\omega)$ that read as:
\begin{equation}
\begin{split}
 G^{(n)}_{ab}(\omega) = \langle \Psi_n | c_a \frac{1}{\omega + i\eta + E_n  - H} c_b^\dagger | \Psi_n \rangle \\ 
+ \langle \Psi_n | c_b^\dagger \frac{1}{\omega + i\eta - E_n  + H} c_a | \Psi_n \rangle \\
= G^{>,(n)}_{ab}(\omega) + G^{<,(n)}_{ab}(\omega) ,
\end{split}
\label{retGF}
\end{equation}
where $\ket{\Psi_n}$ is an eigenstate of Eqn.~\eqref{SIAM} with energy $E_n$, and $\eta$ is the imaginary component of the argument of the green function. Upon tri-diagonalization, the inverse of some operator $\hat O$ reads
\begin{align*}
    \hat{O}^{-1} = 
\begin{pmatrix}
a_0 & b_1 & 0 & 0 & \cdots \\
b_1 & a_1 &b_2 & 0 & \cdots  \\
0 & b_2 & a_2 & b_3 & \cdots\\
0 & 0 & b_3 & a_3 & \cdots\\
\vdots & \vdots & \vdots & \vdots &  \ddots & \\
\end{pmatrix}^{-1}
\end{align*}
and the upper left element is given by
\begin{equation}
    (\hat{O})^{-1}_{00} = 
\frac{1}{a_0 + \frac{b_1^2}{a_1 + \frac{b_2^2}{a_2 + \cdots}}}
\label{continued_fraction}
\end{equation}

In order to compute the diagonal elements of eq. \eqref{retGF}, we use the Lanczos algorithm to tri-diagonalize the many-body Hamiltonian. This algorithm takes as input a linear map $\hat O$ and a normalized vector $\ket{v}$, and return the coefficients $a_i$ and $b_i$ in Eq. \eqref{continued_fraction}. The expectation value $\langle v | \hat O  | v\rangle$ corresponds to the first coefficient $a_0$ in the tri-diagonal matrix. Using $\ket{v_g} = c^{\dagger}_a \ket{\Psi_n}$ and $\ket{v_l} = c_a \ket{\Psi_n}$, we can recover $G^{>,(n)}_{aa}(\omega)$ and $G^{<,(n)}_{ab}(\omega)$.
Off-diagonal elements can be computed by considering linear combinations. For instance, applying $c_i + c_j$ and  $c_i^{\dagger} + c_j^{\dagger}$  on $\ket{\Psi_n}$, we obtain $G_{ii} +G_{jj} + G_{ij} + G_{ji}$. Symmetries of the green function ($G_{ij}(z) = G^*_{ji}(z^*)$ in the general case, or $G_{ij}(z) = G_{ji}(z)$ for a real-valued Hamiltonian, allow to extract the off-diagonal component of interest. The temperature-dependent impurity green function, $\bm G^{imp}(\omega)$, for the inverse temperature $\beta$, is computed as 
\begin{equation}
    \bm G_{imp}(\omega) = \frac{1}{Z} \sum_{n=1}^{n_{low}} e^{-\beta E_n} \bm G_{imp}^{(n)}
\end{equation} with $Z = \sum_n e^{-\beta E_n}$. Finally, the impurity self-energy is computed from Dyson equation as 
\begin{equation}
    \bm\Sigma_{imp}(\omega) =  \bm G_{imp}^0(\omega)^{-1} - \bm G_{imp}^{-1}(\omega),
\end{equation}
with $ \bm G_{imp}^0(\omega)^{-1} = (\omega + i\eta) \mathbb{1} - \bm h_{imp} - \bm \Delta(\omega)$.

\subsection{Computational details}

The one-particle electronic Hamilonian used in the DMFT cycle, was obtained using the Relativistic Spin Polarized toolkit (RSPt)\cite{wills_full-potential_2010}, with the DMFT implementation described in Refs. \cite{DMFT2,DMFT4}.
Both \al and \ga phases of Ce were computed in the face-centered cubic primitive cell, with a cell length parameter of $9.116$ and $9.753$ atomic unit, respectively. The $\epsilon$ phase was computed with the experimentally observed body centered tetragonal primitive structure, with length parameters $a = 5.259$ atomic units and with $c/a = 1.67$. 

In order to avoid having the Coulomb tensor as a free parameter, we used c-RPA values, as computed in Ref. \cite{amadon_screened_2014}. This means that for the \al and \ga phases, we used values for the Coulomb interaction of $U=5.2$ and $U=6.6$, respectively, with $J=0.6$ in both phases \cite{amadon_screened_2014}. For all systems shown, DFT+DMFT calculations were performed in a fully charge self-consistent way within a non-relativistic setup, before an additional single shot relativistic calculation was made: the effect of the inclusion of spin-orbit coupling is mainly to split the central quasi-particle peak into a doublet, while reducing the Kondo temperature, but without impacting the physics, as shown by Bieder and Amadon \cite{amadon-2014}. 
A distance to the real axis ($\eta$ in Eqns. \eqref{retGF} and \eqref{fit}) of \SI{0.02}{Ry} and \SI{0.005}{Ry} was used to respectively fit the hybridization functions and compute real-frequency green functions. The fully localized limit with nominal occupancy \cite{Haule2015} was used to model the double counting. Temperature was set to $116 K$, $300 K$, and $300 K$ in the \al, \ga and $\epsilon$ phases respectively.
In all cases, the hybridization function was fitted with $6$ bath orbitals per correlated orbitals, so that a total of $42$ bath orbitals and 14 correlated orbitals were used.

\section{Data Availability Statement}
\subsection{Code Availability}
The program code for the impurity solver presented in this study is available at \url{https://github.com/bslhrzg/Ciimp}

\section{Acknowledgements}
Valuable discussions with Drs. I. Di Marco and J. Jönsson are acknowledged. \\ 
O.E. acknowledge support from the Wallenberg Initiative Materials Science for Sustainability (WISE) funded by the Knut and Alice Wallenberg Foundation (KAW) and the European Research Council through the ERC Synergy Grant 854843-FASTCORR. O.E. also acknowledges support from STandUPP, eSSENCE, the Swedish Research Council (VR) and the Knut and Alice Wallenberg Foundation (KAW-Scholar program) and NL-ECO: Netherlands Initiative for Energy-Efficient Computing (with project number NWA. 1389.20.140) of the NWA research program. 
The computations/data handling were enabled by resources provided by the National Academic Infrastructure for Supercomputing in Sweden (NAISS), partially funded by the Swedish Research Council.

\section{Authors Contributions}
B.H. developed the configuration interaction with machine learning sampling method together with P.T. and performed all calculations. B.H. wrote the initial version of the paper. O.E. conceived the project and discussed the results. All authors contributed to the final version of the text.

\section{Competing Interests}

The authors declare no competing interests.

\bibliographystyle{naturemag}   
\bibliography{bibli_clean}            

@Article{	  oe1,
  title		= {The α-γ transition in cerium is a Mott transition},
  volume	= {30},
  doi		= {10.1080/14786439808206574},
  journal	= {Philosophical Magazine},
  publisher	= {Informa UK Limited},
  author	= {Johansson, B\"{o}rje},
  year		= {1974},
  pages		= {469–482}
}

@Article{	  oe2,
  title		= {Kondo Volume Collapse and the
		  $\ensuremath{\gamma}\ensuremath{\rightarrow}\ensuremath{\alpha}$
		  Transition in Cerium},
  author	= {Allen, J. W. and Martin, Richard M.},
  journal	= {Phys. Rev. Lett.},
  volume	= {49},
  issue		= {15},
  pages		= {1106--1110},
  numpages	= {0},
  year		= {1982},
  publisher	= {American Physical Society},
  doi		= {10.1103/PhysRevLett.49.1106}
}

@InBook{	  oe3,
  title		= {Chapter 4 Cerium},
  doi		= {10.1016/s0168-1273(78)01008-9},
  booktitle	= {Metals},
  publisher	= {Elsevier},
  author	= {Koskenmaki, David C. and Gschneidner, Karl A.},
  year		= {1978},
  pages		= {337–377}
}

@Article{	  oe6,
  title		= {X-ray Emission Spectroscopy of Cerium Across the
		  $\ensuremath{\gamma}\mathrm{\text{\ensuremath{-}}}\ensuremath{\alpha}$
		  Volume Collapse Transition},
  author	= {Lipp, M. J. and Sorini, A. P. and Bradley, J. and Maddox,
		  B. and Moore, K. T. and Cynn, H. and Devereaux, T. P. and
		  Xiao, Y. and Chow, P. and Evans, W. J.},
  journal	= {Phys. Rev. Lett.},
  volume	= {109},
  issue		= {19},
  pages		= {195705},
  numpages	= {5},
  year		= {2012},
  publisher	= {American Physical Society},
  doi		= {10.1103/PhysRevLett.109.195705}
}

@Article{	  oe8,
  title		= {Cerium Volume Collapse: Results from the Merger of
		  Dynamical Mean-Field Theory and Local Density
		  Approximation},
  volume	= {87},
  doi		= {10.1103/physrevlett.87.276404},
  journal	= {Physical Review Letters},
  publisher	= {American Physical Society (APS)},
  author	= {Held, K. and McMahan, A. K. and Scalettar, R. T.},
  year		= {2001}
}

@Article{	  oe9,
  title		= {Topological phase transition in the archetypal
		  $f$-electron correlated system of cerium},
  author	= {Kim, Junwon and Ryu, Dong-Choon and Kang, Chang-Jong and
		  Kim, Kyoo and Choi, Hongchul and Nam, T.-S. and Min, B.
		  I.},
  journal	= {Phys. Rev. B},
  volume	= {100},
  issue		= {19},
  pages		= {195138},
  numpages	= {6},
  year		= {2019},
  publisher	= {American Physical Society},
  doi		= {10.1103/PhysRevB.100.195138}
}

@Article{	  oe12,
  title		= {$\ensuremath{\alpha}$-$\ensuremath{\gamma}$ transition in
		  cerium: Magnetic form factor and dynamic magnetic
		  susceptibility in dynamical mean-field theory},
  author	= {Chakrabarti, B. and Pezzoli, M. E. and Sordi, G. and
		  Haule, K. and Kotliar, G.},
  journal	= {Phys. Rev. B},
  volume	= {89},
  issue		= {12},
  pages		= {125113},
  numpages	= {4},
  year		= {2014},
  publisher	= {American Physical Society},
  doi		= {10.1103/PhysRevB.89.125113}
}

@Article{	  oe14,
  title		= {Theoretical studies of the high pressure phases in
		  cerium},
  volume	= {67},
  doi		= {10.1103/physrevlett.67.2215},
  journal	= {Physical Review Letters},
  publisher	= {American Physical Society (APS)},
  author	= {Wills, J. M. and Eriksson, Olle and Boring, A. M.},
  year		= {1991},
  pages		= {2215–2218}
}

@Article{	  oe16,
  title		= {First-principles theory for cerium predicts three distinct
		  face-centered cubic phases},
  volume	= {15},
  doi		= {10.1038/s41598-025-03174-6},
  journal	= {Scientific Reports},
  publisher	= {Springer Science and Business Media LLC},
  author	= {S\"{o}derlind, Per and Landa, Alexander and Wu, Christine
		  and Swift, Damian and Johansson, B\"{o}rje},
  year		= {2025}
}

@Article{	  oe17,
  title		= {Electronic correlations in cerium’s high-pressure
		  phases},
  volume	= {30},
  doi		= {10.1088/1361-648x/aadc7c},
  journal	= {Journal of Physics: Condensed Matter},
  publisher	= {IOP Publishing},
  author	= {Lu, Haiyan and Huang, Li},
  year		= {2018},
  pages		= {395601}
}

@Article{	  oe18,
  title		= {A unified picture of the crystal structures of metals},
  volume	= {374},
  doi		= {10.1038/374524a0},
  journal	= {Nature},
  publisher	= {Springer Science and Business Media LLC},
  author	= {S\"{o}derlind, Per and Eriksson, Olle and Johansson,
		  B\"{o}rje and Wills, J. M. and Boring, A. M.},
  year		= {1995},
  pages		= {524–525}
}

@Article{	  dmft1,
  title		= {Dynamical mean-field theory of strongly correlated fermion
		  systems and the limit of infinite dimensions},
  volume	= {68},
  doi		= {10.1103/revmodphys.68.13},
  journal	= {Reviews of Modern Physics},
  publisher	= {American Physical Society (APS)},
  author	= {Georges, Antoine and Kotliar, Gabriel and Krauth, Werner
		  and Rozenberg, Marcelo J.},
  year		= {1996},
  pages		= {13–125}
}

@Article{	  dmft2,
  title		= {Theory of bulk and surface quasiparticle spectra for Fe,
		  Co, and Ni},
  volume	= {76},
  doi		= {10.1103/physrevb.76.035107},
  journal	= {Physical Review B},
  publisher	= {American Physical Society (APS)},
  author	= {Grechnev, Alexei and Di Marco, I. and Katsnelson, M. I.
		  and Lichtenstein, A. I. and Wills, John and Eriksson,
		  Olle},
  year		= {2007}
}

@Article{	  dmft3,
  title		= {Charge self-consistent dynamical mean-field theory based
		  on the full-potential linear muffin-tin orbital method:
		  Methodology and applications},
  volume	= {55},
  doi		= {10.1016/j.commatsci.2011.11.032},
  journal	= {Computational Materials Science},
  publisher	= {Elsevier BV},
  author	= {Grån\"{a}s, O. and Di Marco, I. and Thunstr\"{o}m, P. and
		  Nordstr\"{o}m, L. and Eriksson, O. and Bj\"{o}rkman, T. and
		  Wills, J.M.},
  year		= {2012},
  pages		= {295–302}
}

@Article{	  dmft4,
  title		= {Electronic Entanglement in Late Transition Metal Oxides},
  volume	= {109},
  doi		= {10.1103/physrevlett.109.186401},
  journal	= {Physical Review Letters},
  publisher	= {American Physical Society (APS)},
  author	= {Thunstr\"{o}m, Patrik and Di Marco, Igor and Eriksson,
		  Olle},
  year		= {2012}
}

@Article{	  oe13,
  title		= {Magnetic susceptibility of cerium: An LDA$+$DMFT study},
  author	= {Streltsov, S. V. and Gull, E. and Shorikov, A. O. and
		  Troyer, M. and Anisimov, V. I. and Werner, P.},
  journal	= {Phys. Rev. B},
  volume	= {85},
  issue		= {19},
  pages		= {195109},
  numpages	= {5},
  year		= {2012},
  publisher	= {American Physical Society},
  doi		= {10.1103/PhysRevB.85.195109}
}

@Article{	  huron1973,
  title		= {Iterative perturbation calculations of ground and excited
		  state energies from multiconfigurational zeroth-order
		  wavefunctions},
  volume	= {58},
  doi		= {10.1063/1.1679199},
  journal	= {The Journal of Chemical Physics},
  publisher	= {AIP Publishing},
  author	= {Huron, B. and Malrieu, J. P. and Rancurel, P.},
  year		= {1973},
  pages		= {5745–5759}
}

@Article{	  garniron2019,
  title		= {Quantum Package 2.0: An Open-Source Determinant-Driven
		  Suite of Programs},
  volume	= {15},
  doi		= {10.1021/acs.jctc.9b00176},
  journal	= {Journal of Chemical Theory and Computation},
  publisher	= {American Chemical Society (ACS)},
  author	= {Garniron, Yann and Applencourt, Thomas and Gasperich,
		  Kevin and Benali, Anouar and Ferté, Anthony and Paquier,
		  Julien and Pradines, Barthélémy and Assaraf, Roland and
		  Reinhardt, Peter and Toulouse, Julien and Barbaresco,
		  Pierrette and Renon, Nicolas and David, Grégoire and
		  Malrieu, Jean-Paul and Véril, Mickaël and Caffarel,
		  Michel and Loos, Pierre-Fran\c{c}ois and Giner, Emmanuel
		  and Scemama, Anthony},
  year		= {2019},
  pages		= {3591–3609}
}

@Article{	  amadon_screened_2014,
  title		= {Screened {Coulomb} interaction calculations: {cRPA}
		  implementation and applications to dynamical screening and
		  self-consistency in uranium dioxide and cerium},
  volume	= {89},
  copyright	= {http://link.aps.org/licenses/aps-default-license},
  shorttitle	= {Screened {Coulomb} interaction calculations},
  doi		= {10.1103/PhysRevB.89.125110},
  language	= {en},
  urldate	= {2025-02-12},
  journal	= {Physical Review B},
  author	= {Amadon, Bernard and Applencourt, Thomas and Bruneval,
		  Fabien},
  year		= {2014},
  pages		= {125110}
}

@Article{	  wieliczka_high-resolution_1984,
  title		= {High-resolution photoemission study of γ - and α
		  -cerium},
  volume	= {29},
  copyright	= {http://link.aps.org/licenses/aps-default-license},
  doi		= {10.1103/PhysRevB.29.3028},
  language	= {en},
  urldate	= {2025-02-12},
  journal	= {Physical Review B},
  author	= {Wieliczka, D. M. and Olson, C. G. and Lynch, D. W.},
  year		= {1984},
  pages		= {3028--3030}
}

@Article{	  wuilloud_electronic_1983,
  title		= {Electronic structure of $\gamma$ - and $\alpha$ - {Ce}},
  volume	= {28},
  copyright	= {http://link.aps.org/licenses/aps-default-license},
  doi		= {10.1103/PhysRevB.28.7354},
  language	= {en},
  urldate	= {2025-02-12},
  journal	= {Physical Review B},
  author	= {Wuilloud, E. and Moser, H. R. and Schneider, W. -D. and
		  Baer, Y.},
  year		= {1983},
  pages		= {7354--7357}
}

@Article{	  herper_combining_2017,
  title		= {Combining electronic structure and many-body theory with
		  large databases: {A} method for predicting the nature of 4
		  f states in {Ce} compounds},
  volume	= {1},
  copyright	= {https://link.aps.org/licenses/aps-default-license},
  shorttitle	= {Combining electronic structure and many-body theory with
		  large databases},
  doi		= {10.1103/PhysRevMaterials.1.033802},
  language	= {en},
  urldate	= {2025-02-12},
  journal	= {Physical Review Materials},
  author	= {Herper, H. C. and Ahmed, T. and Wills, J. M. and Di Marco,
		  I. and Björkman, T. and Iuşan, D. and Balatsky, A. V. and
		  Eriksson, O.},
  year		= {2017},
  pages		= {033802}
}

@Book{		  wills_full-potential_2010,
  address	= {Berlin, Heidelberg},
  series	= {Springer {Series} in {Solid}-{State} {Sciences}},
  title		= {Full-{Potential} {Electronic} {Structure} {Method}:
		  {Energy} and {Force} {Calculations} with {Density}
		  {Functional} and {Dynamical} {Mean} {Field} {Theory}},
  volume	= {167},
  copyright	= {https://www.springernature.com/gp/researchers/text-and-data-mining},
  isbn		= {978-3-642-15143-9 978-3-642-15144-6},
  shorttitle	= {Full-{Potential} {Electronic} {Structure} {Method}},
  language	= {en},
  urldate	= {2025-02-12},
  publisher	= {Springer Berlin Heidelberg},
  author	= {Wills, John M. and Eriksson, Olle and Andersson, Per and
		  Delin, Anna and Grechnyev, Oleksiy and Alouani, Mebarek},
  year		= {2010},
  doi		= {10.1007/978-3-642-15144-6}
}

@Article{	  lu_exact_2017,
  title		= {Exact diagonalization as an impurity solver in dynamical
		  mean field theory},
  volume	= {226},
  doi		= {10.1140/epjst/e2017-70042-4},
  language	= {en},
  urldate	= {2025-02-12},
  journal	= {The European Physical Journal Special Topics},
  author	= {Lu, Yi and Haverkort, Maurits W.},
  year		= {2017},
  pages		= {2549--2564}
}

@Article{	  lu_natural-orbital_2019,
  title		= {Natural-orbital impurity solver and projection approach
		  for {Green}'s functions},
  volume	= {100},
  doi		= {10.1103/PhysRevB.100.115134},
  language	= {en},
  urldate	= {2025-02-12},
  journal	= {Physical Review B},
  author	= {Lu, Y. and Cao, X. and Hansmann, P. and Haverkort, M. W.},
  year		= {2019},
  pages		= {115134}
}

@Article{	  lu_efficient_2014,
  title		= {Efficient real-frequency solver for dynamical mean-field
		  theory},
  volume	= {90},
  copyright	= {http://link.aps.org/licenses/aps-default-license},
  doi		= {10.1103/PhysRevB.90.085102},
  language	= {en},
  urldate	= {2025-02-12},
  journal	= {Physical Review B},
  author	= {Lu, Y. and Höppner, M. and Gunnarsson, O. and Haverkort,
		  M. W.},
  year		= {2014},
  pages		= {085102}
}

@Article{	  oe7,
  title		= {The alpha-gamma transition of {Cerium} is entropy-driven},
  volume	= {96},
  doi		= {10.1103/PhysRevLett.96.066402},
  language	= {en},
  urldate	= {2025-02-20},
  journal	= {Physical Review Letters},
  author	= {Amadon, B. and Biermann, S. and Georges, A. and
		  Aryasetiawan, F.},
  year		= {2006},
  note		= {arXiv:cond-mat/0504732},
  pages		= {066402}
}

@Article{	  oe11,
  title		= {Thermodynamics of the α - γ transition in cerium from
		  first principles},
  volume	= {89},
  copyright	= {http://link.aps.org/licenses/aps-default-license},
  doi		= {10.1103/PhysRevB.89.195132},
  language	= {en},
  urldate	= {2025-02-28},
  journal	= {Physical Review B},
  author	= {Bieder, J. and Amadon, B.},
  year		= {2014},
  pages		= {195132}
}

@Article{	  oe15,
  title		= {Theoretical investigation of the high-pressure phases of
		  {Ce}},
  volume	= {57},
  copyright	= {http://link.aps.org/licenses/aps-default-license},
  doi		= {10.1103/PhysRevB.57.2091},
  language	= {en},
  urldate	= {2025-02-28},
  journal	= {Physical Review B},
  author	= {Ravindran, P. and Nordström, L. and Ahuja, R. and Wills,
		  J. M. and Johansson, B. and Eriksson, O.},
  year		= {1998},
  pages		= {2091--2101}
}

@Article{	  oe10,
  title		= {Electronic structure of cerium: {A} comprehensive
		  first-principles study},
  volume	= {99},
  shorttitle	= {Electronic structure of cerium},
  doi		= {10.1103/PhysRevB.99.045122},
  language	= {en},
  urldate	= {2025-03-04},
  journal	= {Physical Review B},
  author	= {Huang, Li and Lu, Haiyan},
  year		= {2019},
  pages		= {045122}
}

@Article{	  oe4,
  title		= {Standard model of the rare earths analyzed from the
		  {Hubbard} {I} approximation},
  volume	= {94},
  copyright	= {http://link.aps.org/licenses/aps-default-license},
  doi		= {10.1103/PhysRevB.94.085137},
  language	= {en},
  urldate	= {2025-03-07},
  journal	= {Physical Review B},
  author	= {Locht, I. L. M. and Kvashnin, Y. O. and Rodrigues, D. C.
		  M. and Pereiro, M. and Bergman, A. and Bergqvist, L. and
		  Lichtenstein, A. I. and Katsnelson, M. I. and Delin, A. and
		  Klautau, A. B. and Johansson, B. and Di Marco, I. and
		  Eriksson, O.},
  year		= {2016},
  pages		= {085137}
}

@Article{	  herzog2023,
  title		= {Solving the Schr\"{o}dinger Equation in the Configuration
		  Space with Generative Machine Learning},
  volume	= {19},
  doi		= {10.1021/acs.jctc.2c01216},
  journal	= {Journal of Chemical Theory and Computation},
  publisher	= {American Chemical Society (ACS)},
  author	= {Herzog, Basile and Casier, Bastien and Lebègue,
		  Sébastien and Rocca, Dario},
  year		= {2023},
  pages		= {2484–2490}
}

@Article{	  sorgenfrei2024,
  title		= {Theory of x-ray absorption spectroscopy for ferrites},
  volume	= {109},
  doi		= {10.1103/physrevb.109.115126},
  journal	= {Physical Review B},
  publisher	= {American Physical Society (APS)},
  author	= {Sorgenfrei, Felix and Alouani, Mébarek and Sch\"{o}tt,
		  Johan and J\"{o}nsson, H. Johan M. and Eriksson, Olle and
		  Thunstr\"{o}m, Patrik},
  year		= {2024}
}

@Article{	  paul2019,
  title		= {Applications of DFT + DMFT in Materials Science},
  volume	= {49},
  doi		= {10.1146/annurev-matsci-070218-121825},
  journal	= {Annual Review of Materials Research},
  publisher	= {Annual Reviews},
  author	= {Paul, Arpita and Birol, Turan},
  year		= {2019},
  pages		= {31–52}
}

@Article{	  zgid2012,
  title		= {Truncated configuration interaction expansions as solvers
		  for correlated quantum impurity models and dynamical
		  mean-field theory},
  volume	= {86},
  doi		= {10.1103/physrevb.86.165128},
  journal	= {Physical Review B},
  publisher	= {American Physical Society (APS)},
  author	= {Zgid, Dominika and Gull, Emanuel and Chan, Garnet
		  Kin-Lic},
  year		= {2012}
}

@Article{	  bilous2025,
  title		= {Neural-network-supported basis optimizer for the
		  configuration interaction problem in quantum many-body
		  clusters: Feasibility study and numerical proof},
  volume	= {111},
  doi		= {10.1103/physrevb.111.035124},
  journal	= {Physical Review B},
  publisher	= {American Physical Society (APS)},
  author	= {Bilous, Pavlo and Thirion, Louis and Menke, Henri and
		  Haverkort, Maurits W. and Pálffy, Adriana and Hansmann,
		  Philipp},
  year		= {2025}
}

@Article{	  kim2020,
  title		= {Alleviating the sign problem in quantum Monte Carlo
		  simulations of spin-orbit-coupled multiorbital Hubbard
		  models},
  volume	= {101},
  doi		= {10.1103/physrevb.101.045108},
  journal	= {Physical Review B},
  publisher	= {American Physical Society (APS)},
  author	= {Kim, Aaram J. and Werner, Philipp and Valentí, Roser},
  year		= {2020}
}

@Article{	  whitten1969,
  title		= {Configuration Interaction Studies of Ground and Excited
		  States of Polyatomic Molecules. I. The CI Formulation and
		  Studies of Formaldehyde},
  volume	= {51},
  doi		= {10.1063/1.1671985},
  journal	= {The Journal of Chemical Physics},
  publisher	= {AIP Publishing},
  author	= {Whitten, J. L. and Hackmeyer, Melvyn},
  year		= {1969},
  pages		= {5584–5596}
}

@Article{	  rueff2006,
  title		= {Probing the
		  $\ensuremath{\gamma}\mathrm{\text{\ensuremath{-}}}\ensuremath{\alpha}$
		  Transition in Bulk Ce under Pressure: A Direct
		  Investigation by Resonant Inelastic X-Ray Scattering},
  author	= {Rueff, J.-P. and Iti\'e, J.-P. and Taguchi, M. and Hague,
		  C. F. and Mariot, J.-M. and Delaunay, R. and Kappler, J.-P.
		  and Jaouen, N.},
  journal	= {Phys. Rev. Lett.},
  volume	= {96},
  issue		= {23},
  pages		= {237403},
  numpages	= {4},
  year		= {2006},
  publisher	= {American Physical Society},
  doi		= {10.1103/PhysRevLett.96.237403}
}

@Article{	  haule2015,
  title		= {Exact Double Counting in Combining the Dynamical Mean
		  Field Theory and the Density Functional Theory},
  volume	= {115},
  doi		= {10.1103/physrevlett.115.196403},
  journal	= {Physical Review Letters},
  publisher	= {American Physical Society (APS)},
  author	= {Haule, Kristjan},
  year		= {2015}
}

@Article{	  amadon-2014,
  title		= {Thermodynamics of the
		  $\ensuremath{\alpha}$-$\ensuremath{\gamma}$ transition in
		  cerium from first principles},
  author	= {Bieder, J. and Amadon, B.},
  journal	= {Phys. Rev. B},
  volume	= {89},
  issue		= {19},
  pages		= {195132},
  numpages	= {7},
  year		= {2014},
  publisher	= {American Physical Society},
  doi		= {10.1103/PhysRevB.89.195132}
}

@PhDThesis{	  vandereb-2000,
  title		= "Cerium, one of nature's purest puzzles",
  author	= "van der Eb, Jeroen Wichert",
  note		= "Relation: http://www.rug.nl/ date\_submitted:2000 Rights:
		  University of Groningen",
  year		= "2000",
  language	= "English",
  isbn		= "9036713226",
  publisher	= "s.n.",
  school	= "Groningen"
}

\clearpage

\section*{Figure legends}
\noindent \textbf{Figure 1. } {Experimental and theoretical spectral function of $\alpha-$Ce for occupied states (left figure) and unoccupied states (right figure). Experimental data are from Ref.\cite{wieliczka_high-resolution_1984}. The intensity difference between unoccupied and occupied states was chosen to represent the occupation of available states.} \\

\noindent \textbf{Figure 2. } {Experimental and theoretical spectral function of $\gamma-$Ce for occupied states (left figure) and unoccupied states (right figure). Experimental data are from Ref.\cite{wuilloud_electronic_1983}. The intensity difference between unoccupied and occupied states was chosen to represent the occupation of available states.} \\

\noindent \textbf{Figure 3. }{$\alpha-$Ce: Comparison of CT-QMC total spectral function \cite{OE10} with this work (upper panel) and the $4f$ projected results (lower panel). For details of the calculations see main text.} \\

\noindent \textbf{Figure 4. }{$\gamma-$Ce: Comparison of CT-QMC total spectral function \cite{OE10} with this work (upper panel) and the $4f$ projected results (lower panel). For details of the calculations see main text.} \\

\noindent \textbf{Figure 5. }{Energy dispersion of $\alpha-$Ce obtained from the same calculation as shown in Fig.\ref{fig:alpha_dos} for k-integrated data.} \\

\noindent \textbf{Figure 6. }{Energy dispersion of $\gamma-$Ce obtained from the same calculation as shown in Fig.\ref{fig:gamma_dos} for k-integrated data.}\\

\noindent \textbf{Figure 7. }{Trace of the hybridization functions with respect to the $4f$ shell for $\alpha$, $\gamma$ and $\epsilon$ cerium, from converged DFT (dashed lines) and DMFT (solid lines) calculations.} \\

\noindent \textbf{Figure 8. }{Projected $4f$ spectral function for $\epsilon-$Ce, using DFT and DFT+DMFT with $U=\SI{2}{\eV}$ and $U=\SI{5}{\eV}$. For details see main text.}\\

\noindent \textbf{Figure 9. }{Energy dispersion of $\epsilon-$Ce obtained from the same DFT calculation as shown in Fig.\ref{fig:epsilon_dos} for k-integrated data.} \\

\noindent \textbf{Figure 10. }{Energy dispersion of $\epsilon-$Ce obtained from the same DFT+DMFT calculation as shown in Fig.\ref{fig:epsilon_dos} for k-integrated data, with U=2 eV.}

\end{document}